\documentclass[10pt,journal,compsoc]{IEEEtran}

\usepackage{graphicx}
\graphicspath{{figure/}}
\usepackage{amsmath,amssymb}
\usepackage{hyperref}
\hypersetup{
  colorlinks=true,
  linkcolor=blue,
  urlcolor=blue,
  citecolor=blue
}
\usepackage{siunitx}
\usepackage{tikz}
\usetikzlibrary{arrows.meta}
\usetikzlibrary{positioning}
\usetikzlibrary{shapes}
\usetikzlibrary{automata}
\usetikzlibrary{calc}

\newcommand{\Prob}[1]{\mathrm{Pr}(#1)}

\onecolumn

\title{Multiverse: A Simulator for Evaluating Entanglement Routing in Quantum Networks}

\author{
    Amar~Abane,
    Junxiao~Shi,
    Van~Sy~Mai,
    Abderrahim~Amlou,
    Abdella~Battou
    \\[0.6em]
    National Institute of Standards and Technology, USA
}

\date{}

\begin{document}

\maketitle

\begin{abstract}\looseness-1We present MQNS, a discrete-event simulator for rapid evaluation of entanglement routing under dynamic, heterogeneous configurations. MQNS supports runtime-configurable purification, swapping, memory management, and routing, within a unified qubit lifecycle and integrated link-architecture models. A modular, minimal design keeps MQNS architecture-agnostic, enabling fair, reproducible comparisons across paradigms and facilitating future emulation.
\end{abstract}

\maketitle

\section{Introduction}
Quantum networks (QNs) enable distributed computing and quantum key distribution by generating fragile long-distance entanglement despite photon loss, decoherence, and noisy operations. End-to-end distribution combines routing, swapping, and purification across heterogeneous hardware (color centers, ions, superconductors). Designing networks that account for diverse hardware processes, encoding schemes, and interdependent operations remains a major challenge.

While different entanglement routing paradigms have been proposed \cite{abane2025entanglementRouting} their evaluation remains limited by the lack of comprehensive, configurable simulation tools. Although several simulators exist, most entanglement routing studies rely on custom single-purpose simulators or analytical models that fail to capture dynamic behaviors or the interplay of repeater functions. As QNs are increasingly viewed as communication systems—drawing on classical networking concepts—simulators must support comparative evaluations of network architectures and routing paradigms.

We introduce Multiverse Quantum Network Simulator (MQNS)\footnote{https://github.com/usnistgov/mqns}, a discrete-event simulator that enables quick evaluation of entanglement distribution under dynamic configurations.
MQNS offers configurable routing algorithms, swapping strategies, purification schemes, and memory management policies. It models a large set of physical link architectures and encoding schemes for Einstein–Podolsky–Rosen (EPR) pairs generation.

The remaining of this article reviews existing simulators and their limitations, introduces the core functions and key conceptual components of MQNS. Implementation aspects and a demonstration architecture are then described, followed by validation results and use cases.

\section{Related Work}
\label{sec:rw} 
SimulaQron \cite{Dahlberg2018simulaqron} provides an emulation-like environment for application-layer testing but lacks built-in routing functionalities. 
While NetSquid \cite{Coopmans2021netsquid} offers quantum state tracking and interfaces for protocol development, it requires extensive customization for advanced routing.
SeQUeNCe \cite{Wu2021sequence} is recognized for its precision and modularity but is limited to static swapping and routing configurations.
QuNetSim \cite{Diadamo2021qunetsim} is designed for small-scale QNs with an OSI-inspired architecture abstracting the network behavior. Consequently, QuNetSim is best suited for evaluating upper-layer quantum applications rather than routing and architectures.
QuISP \cite{Satoh2022quisp} provides scalable modeling with error tracking and introduces the RuleSet protocol for programmable entanglement distribution. However, its reliance on the RuleSet architecture makes dynamic routing dependent on protocol extensions.
SimQN \cite{Chen2023simqn} adopts an architecture-agnostic design, aligning with the current lack of a recognized QN architecture and protocol stack. Its limitation lies in a design that does not accommodate inter-paradigm evaluation and a rigid implementation of critical functions such as swapping and purification.

Each simulator reflects evolving research needs, progressing from application-oriented tools to network-level solutions inspired by classical networking.
While some simulators can be extended to support functions such as routing, swapping, and purification, their configurations are typically static or tied to specific protocol designs. MQNS differs by unifying these functions into a single framework with runtime reconfigurability, supporting diverse routing paradigms and network architectures within the same platform.

\section{Design Approach}
\label{approach}

MQNS targets fast evaluation of QN architectures and network-layer mechanisms. Its minimal design keeps it architecture-agnostic. By avoiding artificial constructs such as global resource managers, it allows realistic architecture and protocol design and can evolve into a lightweight emulation platform. MQNS design is based on the following functions.

Entanglement routing strategies follow two main paradigms: proactive and reactive, both supporting distributed or centralized deployment \cite{abane2025entanglementRouting}.
In proactive routing, paths are computed and installed before the forwarding phase, which includes EPR generation, purification, and swapping.
In reactive routing, nodes continuously attempt elementary EPR generation on all physical channels. Routing is performed on the instant topology formed by generated EPRs, then informs the forwarding operations.

Swapping strategies, in the form of static orders such as sequential or balanced-tree and policies such as \textit{swap-as-soon-as-possible} (swap-asap) should be configurable at runtime through routing instructions.

Purification schemes can also be dynamically configurable, allowing the selection of elementary links or segments—EPRs that span two or more elementary links—to purify for a certain number of rounds.

Memory management policies determine how qubits are assigned to physical channels at topology creation. 

Multiplexing schemes define how qubits are allocated to paths and which qubits are selected for operations when multiple candidates are available.

\section{Key Components}
\label{components}
MQNS is structured around four conceptual components that can be customized and composed to realize diverse networking scenarios.

\subsection{Time Scheduling}
\begin{figure}
\centerline{\includegraphics[width=25pc]{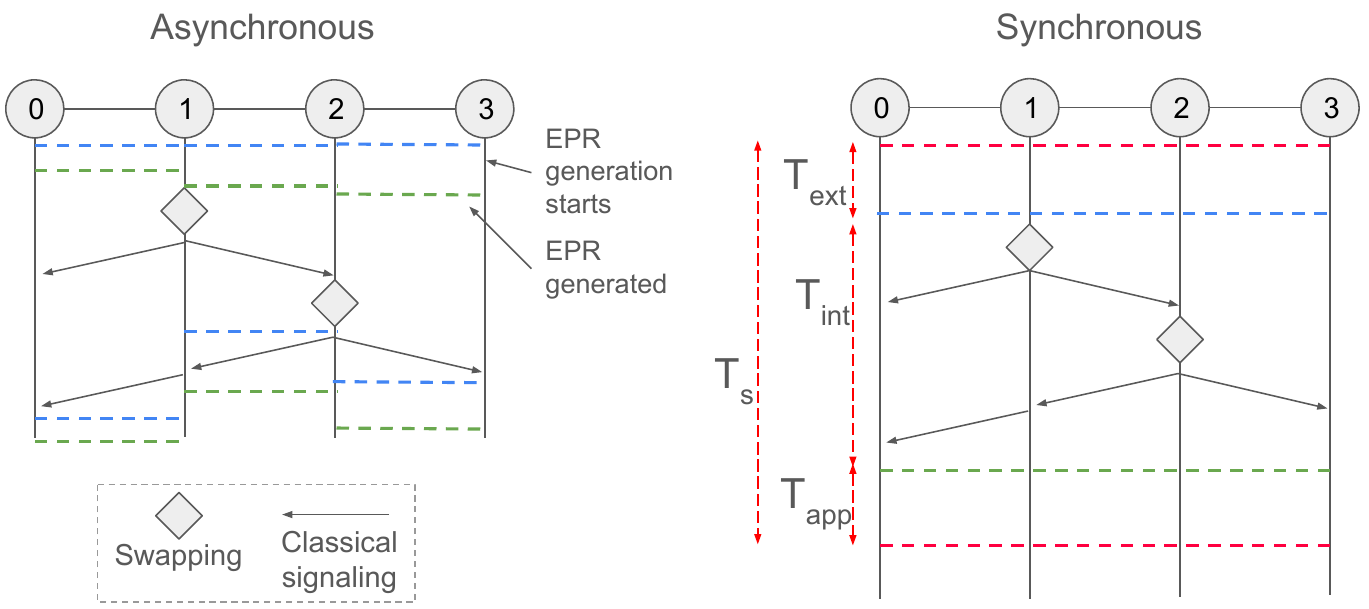}}
\caption{Timing modes in entanglement routing. Synchronous (left) and Asynchronous (right)}
\label{fig:scheduling}
\end{figure}

Entanglement distribution can be coordinated in an asynchronous or synchronous (slotted) timing mode. As illustrated in Figure \ref{fig:scheduling}, asynchronous EPR generation can start as soon as two neighboring nodes have a pair of available qubits. Swapping and purification can take place anytime relevant EPRs are available, in accordance with the routing instructions.
In synchronous mode, each time slot $T_s$ is divided into three phases: an external phase $T_{ext}$ for EPR generation, an internal phase $T_{int}$ for entanglement swapping and purification, and an application phase $T_{app}$ for users to consume E2E entanglements.
In reactive routing under synchronous mode, a dedicated routing interval $T_r$ is introduced after $T_{ext}$ to collect the instant entanglement topology and retrieve routing instructions (e.g., from a controller) before the internal phase begins.

\subsection{Memory Management}
Memory management supports entanglement distribution needs from static and predefined to dynamic and adaptive configurations.

Qubits can be assigned to channels at topology creation, either in a symmetric way—assigning the same number of qubits to both ends of a channel—or asymmetric.

A two-way reservation exchange is performed between each pair of neighbors prior to EPR generation. This process is initiated by the primary node (defined as the left side of the channel) once a free qubit becomes available, typically after a qubit is consumed or decoheres, and completed when the secondary node confirms availability on its end. This process is immediately followed by the EPR generation.

At runtime, each node uses routing instructions to pre-allocate qubits to installed paths either exclusively (blocking multiplexing) or by splitting qubits across multiple paths (buffer-space multiplexing). 
Alternatively, paths may be installed without predefined qubit allocations, allowing nodes to manage their local qubits autonomously. 
In such cases, statistical multiplexing assigns a newly generated EPR to the first available EPR waiting for swapping.

\subsection{Qubit Lifecycle}
\begin{figure*}[t]
\centering
\begin{tikzpicture}[
    scale=0.75, transform shape,
    >=Stealth,
    node distance=2.7cm and 2.0cm,
    every node/.style={font=\small, align=center},
    state/.style={
        draw,
        rounded corners,
        minimum width=1.8cm,
        minimum height=0.9cm,
        align=center
    },
    note/.style={font=\scriptsize, align=left, text width=4.5cm}
]

\node[state, double, double distance=1pt] (raw) {RAW};
\node[state,right=2.5cm of raw]           (active)   {ACTIVE};
\node[state,right=3.4cm of active]        (reserved) {RESERVED};
\node[state,right=2.6cm of reserved]      (ent)      {ENTANGLED};
\node[state,right=3cm of ent]           (purif)    {PURIF};

\node[state,below=5cm of active]     (eligible) {ELIGIBLE};
\node[state,right= 5cm of eligible]      (release)  {RELEASE};
\node[state,right=3cm of release]         (pending)  {PENDING};

\path[->]
  (raw)      edge node[below=-0.38]{start reservation \\(wait for available \\remote qubit)}                        (active)
  (active)   edge node[below=-0.38]{remote qubit available \\ (partner node\\ responded)} (reserved)

  (reserved) edge[left=5] node[above]{EPR pair created}            (ent)
  (reserved) edge[loop above] node{simulate EPR attempts}               ()

  (ent)      edge node[above]{swap conditions met \\ (node's turn to swap)}      (purif)
  (ent)      edge[bend left=35]  node[below,pos=0.6]{Remote swap \\ or purification failed} (release)

  (purif)    edge[bend left=30] node[above,pos=0.2]{start round\\(request sent)} (pending)
  (pending)  edge[bend left=30] node[below,pos=0.45]{round success} (purif)
  (pending)  edge[left=55] node[below,pos=0.5]{round failure} (release)

  (purif)    edge[bend right=15] node[below,pos=0.65]{purification rounds done \\ or purification not needed \\ (EPR can be swapped)} (eligible)

  (eligible) edge node[below=-0.78]{swap / delivery\\consumes qubit \\ If swap, \\ a \texttt{SWAP\_UPDATE} \\ is sent to neighbors.}         (release)
  (release)  edge[bend left=10] node[below,pos=0.8]{qubit freed} (raw);

\node[note, above right=0.5cm and -2.5cm of raw] (rawnote)
  {At node startup, the qubit is in the RAW state. The \textbf{first} reservation may be initiated when a channel becomes active due to a newly installed path (proactive), or during each $T_{ext}$ phase (reactive).};

\node[note, above right=0.5cm and -2.5cm of ent] (entnote)
  {In the ENTANGLED state, the qubit may wait for \texttt{SWAP\_UPDATE}
   or \texttt{PURIF\_SOLICIT}.\\
   Reactive routing may be triggered.};

\node[note, above right=0.5cm and -2.5cm of purif] (purifenote)
  {In the PURIF state, the \\ node verifies purification \\ conditions before swapping. \\ Qubit may wait for EPRs \\ with the same partner.};

\node[note, below=0.1cm of eligible] (eligiblenote)
  {In the ELIGIBLE state, the qubit may wait for an eligible EPR, depending\\
   on multiplexing / routing.\\
   At an end node, the EPR can be delivered to the application.};

\node[note, below right=0.2cm and -2cm of release] (releasenote)
  {In the RELEASE state, the qubit is freed after it has been
   consumed and returns to RAW, where a new qubit reservation can be triggered to generate a new EPR.};

\node[note, below right=0.2cm and -1.6cm of pending] (pendingenote)
  {In the PENDING state, the qubit waits for purification response from partner node.};

\end{tikzpicture}
\caption{Qubit state machine}
\label{fig:qubit-state-machine}
\end{figure*}
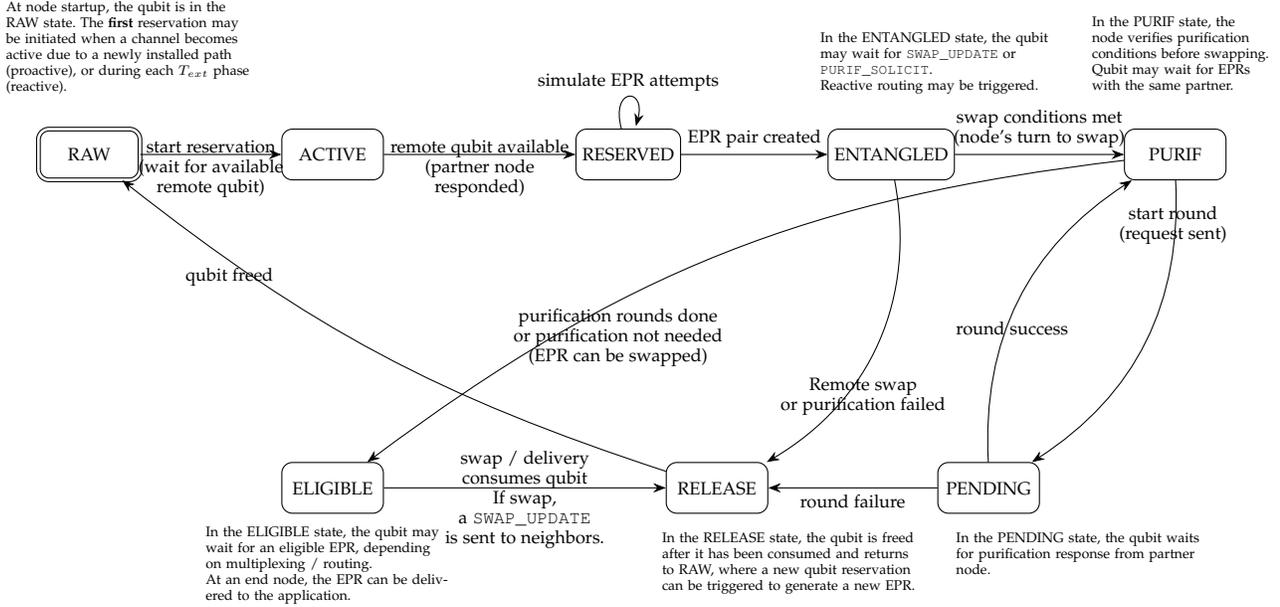

Qubits follow a finite state machine\cite{Bacciottini2023Redip} depicted in Figure \ref{fig:qubit-state-machine}.
Each qubit begins in the \texttt{RAW} state and transitions to the \texttt{ACTIVE} state during the qubit reservation process. When two qubits are reserved for EPR generation, they enter the \texttt{RESERVED} state.
Once an EPR is generated, both qubits move to the \texttt{ENTANGLED} state. 

A qubit in the \texttt{PURIF} state means that its EPR satisfies swapping conditions but has to apply purification configuration first. If purification is required, the qubit undergoes the number of specified rounds looping between \texttt{PURIF} and \texttt{PENDING} states. When the purification requirements are satisfied, the qubit advances to the \texttt{ELIGIBLE} state. If purification fails during the process, the qubits are released.
Purification is coordinated using classical message exchange.

Multiplexing is integrated at the \texttt{ELIGIBLE} stage. In blocking and buffer-space schemes, a qubit in the \texttt{ELIGIBLE} state is swapped only with another local qubit allocated to the same path.
In statistical multiplexing, an \texttt{ELIGIBLE} qubit may be swapped with any other entangled qubit at the same node.

\subsection{Entanglement Link Architectures}
Several link architectures exist for entangling remote matter qubits via photons. We model these architectures based on quantum hardware and qubit platforms by capturing the main archetypes discussed in the recent literature. \cite{beukers2024remote} \cite{dhara2023entangling}
The models presented below are simplified and can be extended to incorporate more detailed physical parameters or link architectures.

We assume the matter-matter Bell state to be generated between the link endpoints $A$ and $B$ is:
\begin{equation}
\Psi^+_{A,B} = \frac{1}{\sqrt{2}} \left( |0_{A} 1_{B}\rangle + |1_{A} 0_{B}\rangle \right)
\label{qe:psi_AB}
\end{equation}

\textbf{Detection-in-Midpoint (DiM):} Both endpoints generate a matter-photon entangled state locally and emit the photon toward a central Bell State Measurement (BSM) station. The BSM performs a linear-optics-based projection of the two incoming photons into a Bell basis. A successful photon detection pattern heralds the entanglement of the two distant matter qubits. A classical message is sent from the BSM to both endpoints to deliver the result.

The initial qubit-photon state at each endpoint is:
\begin{equation}
\Psi_{m,p} = \sqrt{\alpha} |0_m1_p\rangle + \sqrt{1-\alpha} |1_m0_p\rangle
\label{eq:psi_local}
\end{equation}

where $m$ and $p$ subscripts correspond to the matter and photon states respectively, and $\alpha$ is the superposition initialization parameter of the local qubit.

Single rail encoding is sensitive to photon loss, leading to false-positive heralding where the generated Bell state is mixed with the vacuum state.

To overcome false-positive events, the Barrett-Kok (BK) protocol uses a double-round scheme with a flip of the local qubits between rounds, effectively eliminating the noisy term $|00\rangle\langle00|$. 
Its success probability is:
\begin{equation}
p_{\text{DiM}}^{(BK)} = 2\alpha^2 \eta_{Ab}^2
\label{eq:dim_BK}
\end{equation}

Here, $\eta_{Ab}$ denotes the transmissivity of the quantum channel from $A$ to the BSM $b$. It accounts for both fiber attenuation and the detection efficiency at the BSM:

\begin{equation}
\eta_{Ab} = \eta_{b} \, e^{-L_{A,B}/2L_0}
\end{equation}

where $L_{A,B}$ is the channel length and $L_0$ is the fiber attenuation length.
For simplicity, we assume $\eta_{Ab} = \eta_{Bb}$.

In dual-rail encoding, two photons must be emitted, transmitted, and detected, leading to the success probability:
\begin{equation}
p_{\text{DiM}}^{(D)} = 2\alpha(1 - \alpha)\eta_{Ab}^2.
\label{eq:dim_D}
\end{equation}

\textbf{Sender-Receiver (SR):} One endpoint (i.e., sender) emits a photon entangled with its local qubit. The photon travels to the other endpoint (i.e., receiver), where it interacts with its matter qubit via a coherent gate.
The receiver heralds the successful entanglement and informs the sender via a one-way classical message.
This design typically requires dual-rail encoding to distinguish between successful and failed matter-photon interactions.

The heralding probability is \cite{shapourian2025quantum}:
\begin{equation}
p_{\text{SR}}^{(D)} = \eta_d \, e^{-L_{A,B}/L_0}
\label{eq:sr_D}
\end{equation}
where $\eta_d$ captures receiver-side photon detection and interaction efficiency.

\textbf{Source-in-Midpoint (SiM):} An entangled photon pair source (EPPS) is placed in the middle of the link. Each photon is sent to one endpoint, where it interacts with its local qubit.
A two-way classical exchange between the endpoints is used to herald entanglement creation.

The heralding probability is:
\begin{equation}
p_{\text{SiM}}^{(D)} = \eta_{As}^2
\label{eq:sim_D}
\end{equation}

Where $\eta_{As}$ is the transmissivity of the quantum channel from the source to either endpoint (assumed equal for simplicity) and includes photon propagation loss, interaction, and detection efficiency:
\begin{equation}
\eta_{As} = \eta_{d} \, e^{-L_{A,B} / 2L_0}
\label{qe:eta_rs}
\end{equation}

\textbf{Quantum Transduction:} Link architectures can incorporate quantum transduction to bridge the spectral gap between microwave and optical domains \cite{caleffi2025quantum}.

In DiM, entanglement can be established via electro-optic or optomechanical transducers that emit hybrid microwave-optical photon pairs. At each endpoint, the microwave photon interacts with the local qubit, while the optical photon is sent through fiber to the BSM.

The hybrid microwave-optical photon entanglement has the same form as Eq. \eqref{eq:psi_local}, with $m$ and $p$ corresponding to the photons in the microwave and optical domains, respectively, and $\alpha$ interpreted as the transduction efficiency. Hence, the heralded state and probability are similar in form to the DiM links with BK protocol \eqref{eq:dim_D} or dual-rail \eqref{eq:dim_D} encoding.

A direct transduction of photons between the microwave (or near-infrared/visible) and telecom domains is also possible, although converting photons directly from the microwave to the optical domain remains more challenging than generating hybrid entangled photons.

In the SR architecture, photons emitted by the sender are upconverted to the telecom band for transmission and subsequently downconverted at the receiver to match the appropriate local wavelength.
In the SiM architecture, photons are emitted directly at telecom wavelengths and then downconverted to the relevant domain at both endpoints.
In both scenarios, the heralding probability is modified to include the transduction efficiencies at the photon emitter $\eta_s$ and/or receiver $\eta_r$.

\textbf{Noise Modeling:}
Single-rail encoding using the BK protocol and dual-rail encoding are resilient to photon loss, allowing heralded entangled states to reach unit fidelity—assuming no other noise sources.

In practice, imperfections in sources, gates, and quantum memories introduce additional noise. This is commonly modeled as a depolarizing channel, which transforms the ideal Bell state into a Werner state:
\begin{equation}
\rho = W\,|\Psi^+\rangle\langle\Psi^+| + \frac{1 - W}{4} I
\end{equation}
where $W$ is the Werner parameter and the fidelity is $F = (3W + 1) / 4$.
The simulator uses this model to represent EPR pairs with configurable fidelity.

\textbf{Attempt Duration and Rate:} EPR generation attempts are made independently for each reserved qubit pair. We define the \textit{round} duration as the time from the start of an attempt to when nodes are ready to begin the next. 
The round duration depends on the link configuration \cite{Jones2016DesignAndAnalysis}:

\begin{align*}
\tau^{(D)}_{\text{DiM}} &= \tau_l + \tau_0 \\
\tau^{(BK)}_{\text{DiM}} &= 2 (\tau_l + \tau_0) \\
\tau^{(D)}_{\text{SR}}  &= 2\tau_l + \tau_0 \\
\tau^{(D)}_{\text{SiM}} &= \tau_l + \tau_0 \\
\end{align*}
where $\tau_l$ is the fiber propagation delay and $\tau_0$ is the local operation latency.

To avoid simulating individual entanglement attempts, the simulator uses the link's success probability to sample a geometric random variable representing the trial on which the first success occurs. It then advances the simulation time by the total duration of those attempts, executing only the successful one.

\section{Implementation}
\label{implementation}
\subsection{Software Modules}
Although MQNS does not assume any reference QN architecture, its implementation requires a minimal organization of core modules and entities. We built upon SimQN v0.1.5 due to its lightweight structure, which provided a clean foundation for redesigning and introducing new components.

\begin{figure}
\centerline{\includegraphics[width=20pc]{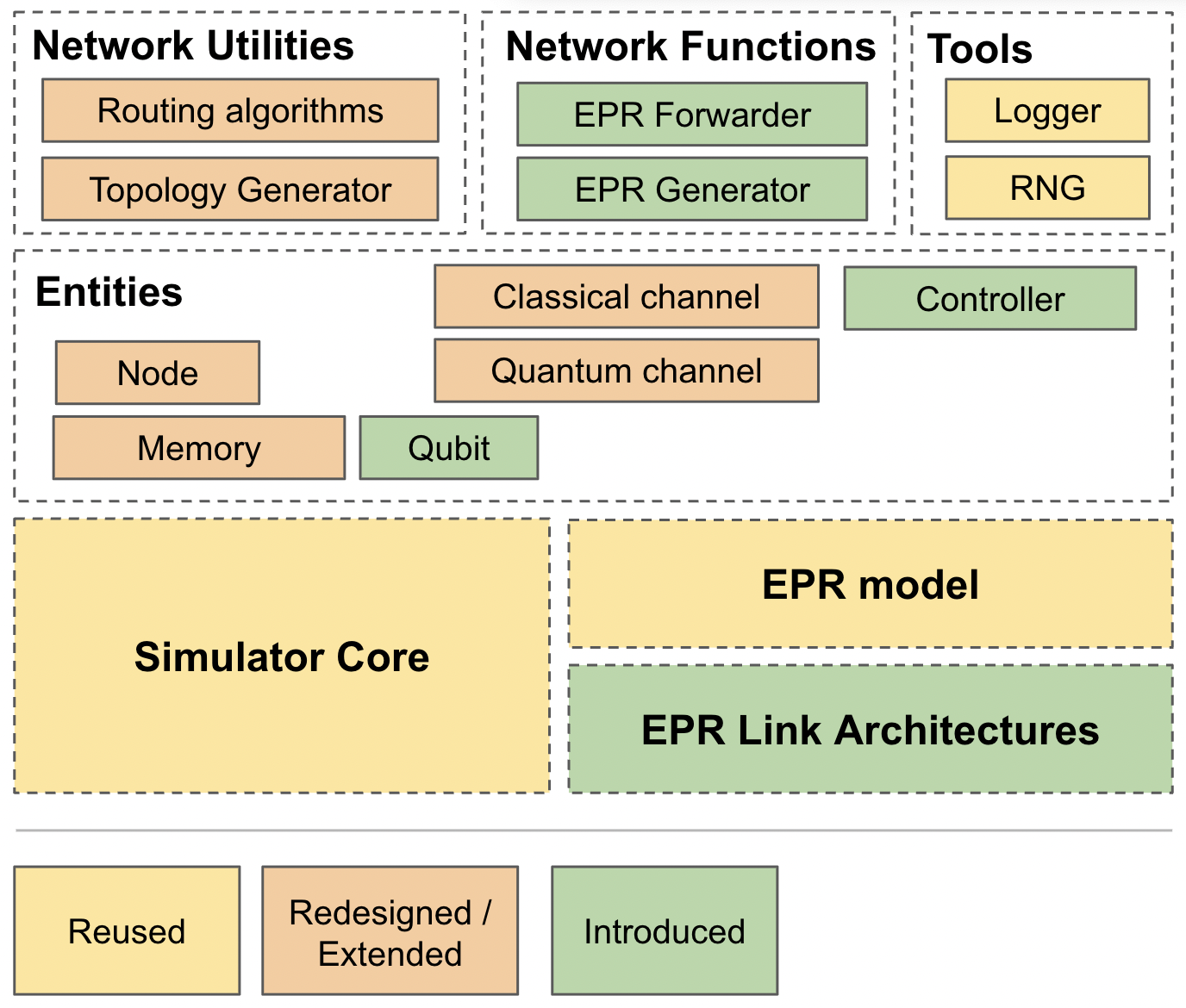}}
\caption{Modules of the simulator, categorized by origin—reused (yellow), redesigned/extended (orange), and newly introduced (green).}
\label{fig:modules}
\end{figure}

Figure \ref{fig:modules} illustrates the modular organization of MQNS and indicates the origin of each component. 
We introduced several key modules, including the EPR generator and forwarder, central controller, and qubit lifecycle manager.

\subsection{Demonstration Network}
We demonstrate a proactive centralized architecture, a commonly assumed deployment where a controller computes paths and installs per-path instructions. Implementation supports synchronous/asynchronous timing, multi-request multipath routing, runtime-configurable swapping orders/policies and purification, and blocking/buffer-space/statistical multiplexing.

Although the demonstration focuses on the proactive centralized mode, the same framework accommodates reactive routing. In this case, the set of qubits that reach the \texttt{ENTANGLED} state at the end of the $T_{ext}$ phase forms the instant logical topology, which informs routing decisions before forwarding operations are initiated. The same forwarding functions (EPR generation, purification, and swapping) are used in the reactive mode. These functions can also operate under a distributed configuration, where the centralized controller is replaced with local control modules instantiated at each node. This require no changes to the underlying abstractions and conceptual components.

\section{Evaluation}
\label{evaluation}
The evaluation is organized to first establish the simulator's validity, then demonstrate its built-in capabilities through two use cases\footnote{Additional use cases are presented in the Supplemental Materials}.
Results are averaged over 1,000 runs \and standard deviations are included in all figures.

\subsection{Primitives Validation}
We validated MQNS against SeQUeNCe, by updating the attempt duration to reflect SeQUeNCe's implementation.

\begin{figure*}
\centerline{\includegraphics[width=\linewidth]{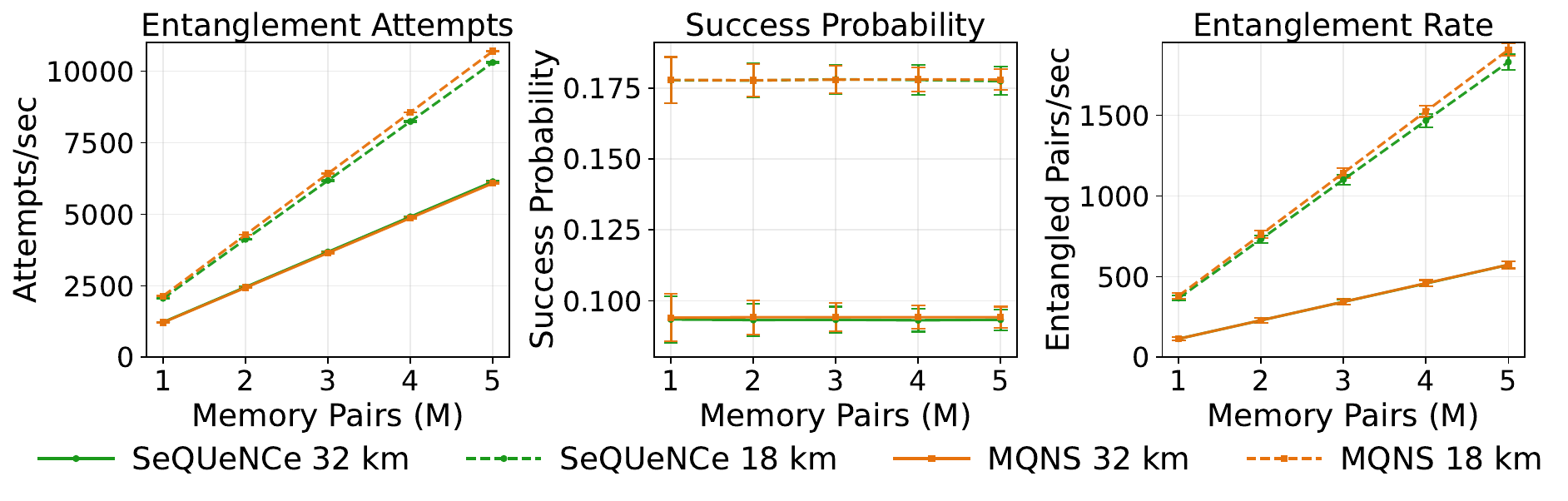}}
\caption{EPR generation performance over isolated links.}
\label{fig:valid_attempts}
\end{figure*}

We first simulated EPR generation on two links with lengths of 32 km and 18 km, respectively. For each link, we varied the number of memory pairs from one to five and simulated each configuration for one second.
The results reported in Figure \ref{fig:valid_attempts} closely mirrored those of SeQUeNCe, especially for success probabilities, with a minor overestimation in the number of entanglement attempts and rates.
Entanglement attempts and rates show limited variability. This behavior is also consistent with SeQUeNCe, where abstraction of certain link details reduces fluctuations in these scenarios.
MQNS produced these results without single-photon attempts, which significantly reduces the number of simulated events.

The attempt rate and entanglement rate generally scale linearly with the number of memory pairs per link in this regime (i.e., high count rates detectors, high frequencies of memories and channels). As expected, longer links exhibit lower success probabilities and reduced attempt rates compared to shorter links.

\begin{figure}
\centering
\includegraphics[width=0.6\linewidth]{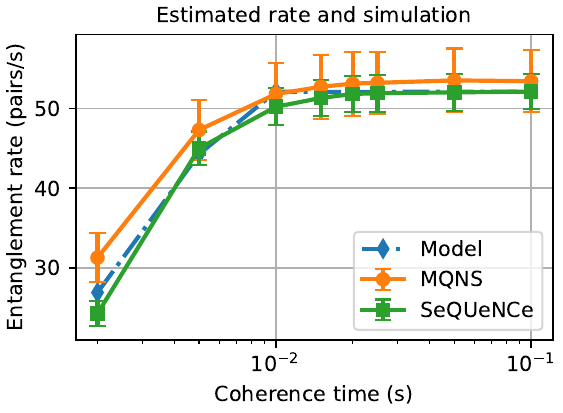}
\caption{E2E entanglement rate for a three-node path.}
\label{fig:valid_thruput}
\end{figure}

We then simulated a three-node linear path with one swapping between the two links used in the previous simulation. Each link was assigned a single memory pair. We varied memory coherence times from 2 ms to 100 ms.
We verified our simulation results using a theoretical model \cite{mai2025optimalorders} (see Supplemental Material), which characterizes the entanglement probability distribution when swapping two EPRs, while accounting for the coherence time in memory. Figure~\ref{fig:valid_thruput} shows a strong alignment between our simulator's output, SeQUeNCe's output and the model's prediction.
At higher coherence times, results converged tightly. At $T_{coh} = 0.002 \, \text{s}$, MQNS overestimated the rate, likely due to slight differences in swapping signaling. However, the general performance trend is similar.

Extending validation to larger networks is difficult due to the lack of reference implementations and implicit simulator differences such as timing assumptions and signaling. This challenge also appears in recent cross-validation work \cite{Chung2025quisp_sequence}, which likewise restricted tests to a single link architecture and one swap. Moreover, many generally available simulators do not natively support features central to our study (e.g., multiplexing strategies, controlled swapping order), and implementing them would entail substantial, non-trivial engineering effort. To keep validation credible and comparable, we focused on small topologies and single-swap scenarios. Scaling further is constrained by the absence of standardized protocol baselines, assumptions divergence, and risk of misinterpreting tool artifacts as protocol effects. These controlled cases provide reliable, reproducible results without overstating generality.

\subsection{Use Case 1: Memory Management}
We propose to investigate how memory allocation affects entanglement distribution in a three-node topology with different link architectures. 
The setup consists of two fiber links of 30 km each. Each node has six qubits. We consider several allocation profiles at the central node, where the number of qubits assigned to each link varies, without exceeding six.
Each allocation profile was simulated under memory coherence times of 5 ms and 10 ms.

\begin{figure}
\centerline{\includegraphics[width=25pc]{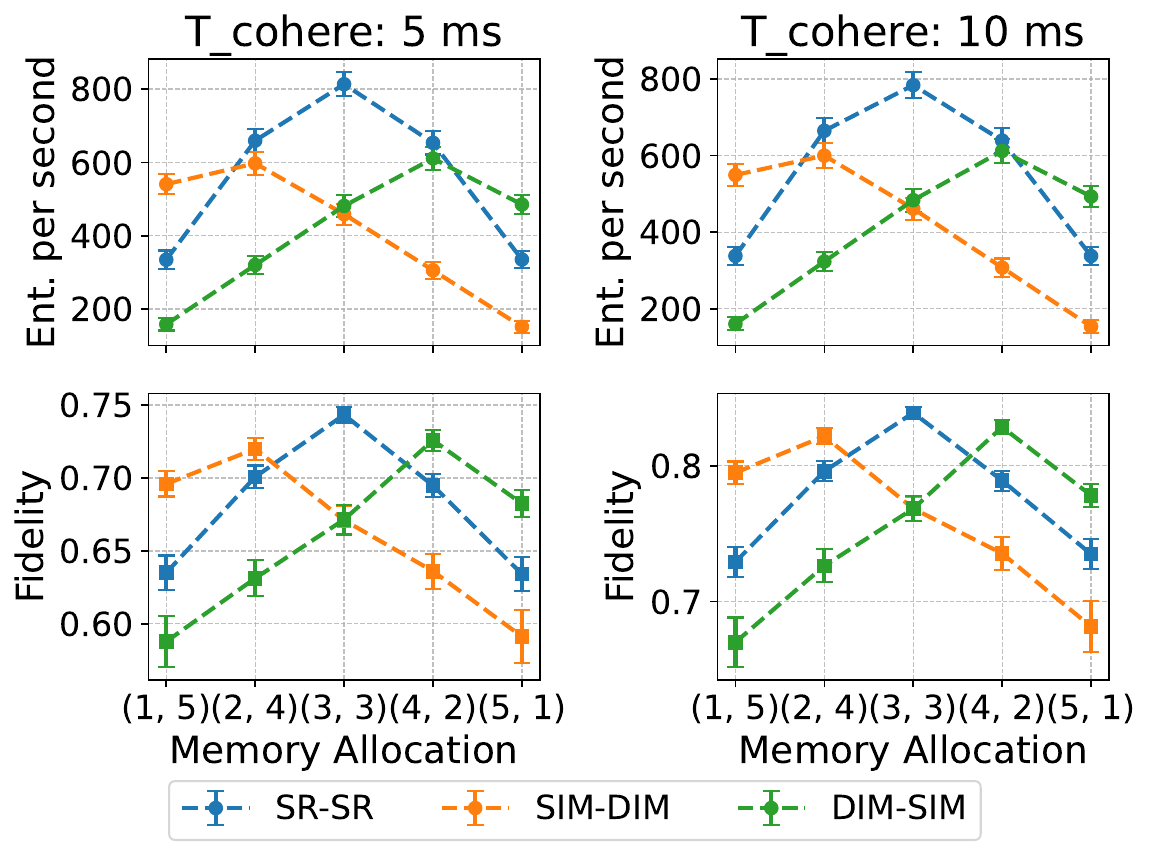}}
\caption{Entanglement throughput and fidelity versus memory allocation and coherence time. (X,Y) labels in the x-axis designate the capacity of left and right links respectively.}
\label{fig:eval_memo_alloc}
\end{figure}

Generation rate and fidelity are shown in Figure \ref{fig:eval_memo_alloc}. Allocations that assign more qubits to link architectures delivering lower success rates tend to yield higher entanglement rates. This is because the link with a lower entanglement success probability becomes the bottleneck unless it is given sufficient resources. When more qubits are allocated to this link, the generation rate increases, which in turn reduces the waiting time for entanglement swapping.

Even in a minimal setup, the way in which qubits are distributed between channels can have a strong impact on overall performance. This simple simulation quickly informs on how qubits need to be allocated along a path for optimal performance, and may also be leveraged to design and evaluate link and node placement in a network.

\subsection{Use Case 2: Swapping Strategies}
We evaluate entanglement swapping strategies combined with memory allocation in a linear six-node network: S–R1–R2–R3–R4–D, with link lengths of 32 km, 18 km, 35 km, 16 km, and 24 km. Each node has six quantum memories.

We compare two memory allocation schemes: a uniform configuration $[3,3,3,3,3]$ and a non-uniform one $[4,2,4,2,4]$, representing the number of qubits assigned to each channel. Coherence times are set to 5 ms, 10 ms, or 20 ms. 
We test five swapping strategies:
\begin{itemize}
\item \textit{asap}: swap immediately as EPRs become available
\item \textit{baln}: swap R1 and R3 first, then R2, then R4
\item \textit{baln2}: swap R2 and R4 first, then R3, then R1
\item \textit{l2r}: swap sequentially from R1 to R4
\item \textit{r2l}: swap sequentially from R4 to R1
\end{itemize}

\begin{figure}
\centerline{\includegraphics[width=\linewidth]{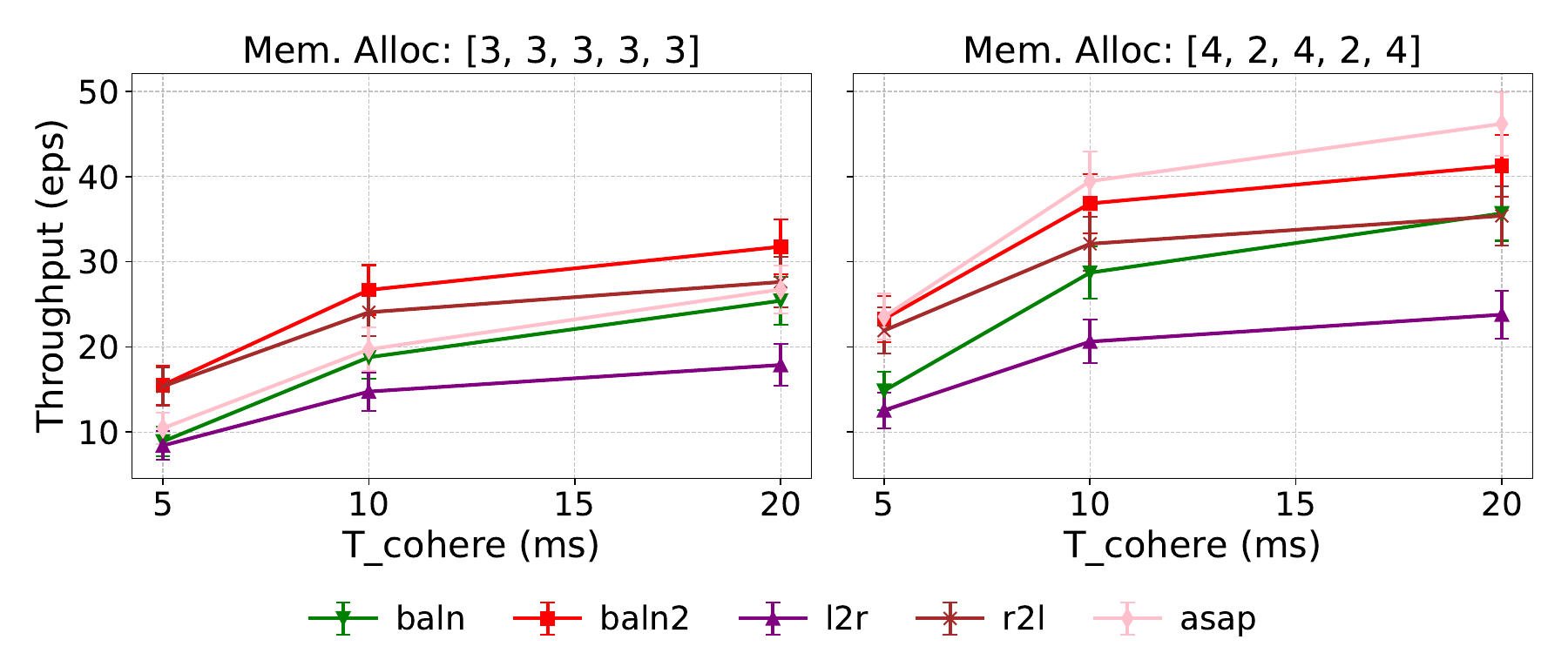}}
\caption{Entanglement throughput with various swapping policies and memory allocation.}
\label{fig:eval_swapping}
\end{figure}

Results are reported in Figure \ref{fig:eval_swapping}. In the uniform configuration, \textit{baln2} achieves the highest throughput at all coherence times, followed closely by \textit{r2l} and \textit{asap}. 
In contrast, \textit{l2r} and \textit{baln} exhibit reduced performance, especially at shorter coherence times.
With non-uniform memory allocation, all strategies show improved performance, with \textit{asap} gaining the most. 
As discussed, longer-distance links often become performance bottlenecks when memory allocation and swapping order are not carefully optimized. Figure \ref{fig:eval_swapping} shows significant throughput improvements as we shift more memories to longer links together with having a more effective swapping order; see recent work \cite{mai2025optimalorders} for further details on how to optimize both in a quantum path.

\section{Conclusion}
\label{conclusion}
Our experience with MQNS surfaced open questions—e.g., how to co-optimize multiplexing and swapping, and how to decide qubit swap-eligibility when multiple paths coexist. MQNS enables these investigations, helping expose bottlenecks and guiding more effective protocols.
Its architecture-agnostic design proved enabling: modeling EPR generation, memory management, and forwarding clarified key repeater operations and interfaces—how qubits/EPRs are addressed and managed, how forwarders coordinate with generators and memories, and how purification and swapping are orchestrated in time and space under dynamic routing. 
This iterative build-and-abstract process positions MQNS not only as a performance tool but also as a vehicle for architectural discovery.

A limitation of the present release is that heterogeneity is modeled only at the link level, with device-level variation handled indirectly via configurable link parameters. Extending the framework with dedicated hardware-specific modules is part of our ongoing work.
Next, we will add reactive and distributed routing and deploy MQNS as an emulation platform for control/management-plane research.

\section{Disclaimer} 
\label{sec:DISCLAIMER}
Any mention of commercial products or references to commercial organizations is for information only; it does not imply recommendation or endorsement by NIST, nor does it imply that the products mentioned are necessarily the best available for the purpose.

\clearpage

\begin{center}
\vspace{2em}
{\LARGE\bfseries Supplementary Materials\par}
\vspace{1em}
\end{center}

\setcounter{section}{0}

\section{A Model for Entanglement Swapping}
We briefly discuss here the model in \cite{mai2025optimalorders} for describing the outcome of swapping two link-level entanglement distributions. We will also introduce a small improvement in the rounding link capacity step in \cite{mai2025optimalorders} using Monte Carlo (MC) simulation approach.

For estimating the entanglement rate, let us assume that time is slotted with each interval of $T_s$, which we will choose later. Assume also that each time slot is partitioned into two phases, one for generating link entanglements and the other for performing swaps and heralding results to end-nodes, which will be denoted by $T_{\rm gen}$ and $T_{\rm her}$, respectively. Then we model the number of entanglements generated over each link $i$ in a time slot as a random variable (RV) $E_i$ following a Binomial distribution
$$
E_i \sim \mathcal{B}(c_i, p_i), \quad i=1,2,
$$
where $c_i$ is the number of attempts over link $i$ within a time slot and $p_i$ the success probability of each corresponding attempt. Clearly, given $E_1$ and $E_2$ entanglements, there is a maximum number of $\min_i\{E_i\}$ possible swaps, each of which succeeds with probability $q$. Let $E_{12}$ denote the number of E2E entanglements, which, when conditioning on link entanglements, can be modeled as another Binomial random variable:
$$
E_{12}|E_1, E_2 \sim \mathcal{B}(\min\{E_1, E_2\}, q).
$$
This allows us to find the probability distribution of $E_{12}$. Specifically, $\Prob{E_{12}=k}$ is simply
\[
\sum_{i=k}^{c_{1}}\sum_{j=k}^{c_{2}}  \Prob{E_1=i} \Prob{E_2=j} {\min\{i,j\} \choose k}
    q^k (1{-}q)^{\min\{i,j\}-k}
\]
for $k=1,\ldots, c_{12}$ and $ \Prob{E_{12}=0} = 1-\sum_{k=1}^{c_{12}} \Prob{E_{12}=k}$ with $c_{12}=\min\{c_{1}, c_2\}$.
As a result, we can estimate the probability distribution of E2E entanglements for any path given any fixed swapping order. Here we restrict ourselves to a three-node path and swapping happens only at the middle node, and the entanglement throughput can be estimated as
\begin{equation}
\frac{\mathbb{E}[E_{12}]}{T_s} = \frac{1}{T_s}\sum_{k=1}^{c_{12}} k\times \Prob{E_{12}=k}. \label{thruput}
\end{equation}

To use the above model,
it remains to select a suitable time slot $T_s$ and link capacities $c_1$ and $c_2$, taking into account the coherence time $T_{\rm cohere}$ of quantum memory.  To this end, we assume that basic link characteristics are available as shown in Figure~4 of the main paper, including the attempt rate, success probability of each attempt, and the link entanglement rate, denoted by $A_i$, $p_i$, and $R_i$ respectively.

Let $T_i$ be the random variable representing the time each link takes to generate one new entanglement (from start or after a swap). Note that the expected value $\mathbb{E}[T_i]$ can be estimated from link characterization step, namely
\[
\mathbb{E}[T_i] \approx 1/R_i.
\]

Clearly, a swap is only possible if there are entanglements on both links and they must be created within a certain time duration so that the E2E entanglement, after being heralded, is still within the coherence time of quantum memory to be usable. Let us denote this period by $T_{\rm cutoff}$, which depends on the coherence time of memory, the delay in heralding swapping results to the end nodes $T_{\rm her} = \max\{L_i\}/c_0$ (assuming classical communications use the same quantum channels with length $L_i$ whose speed of light is $c_0$) and the delay in consuming the end-to-end entanglement by applications $T_{\rm app}$ if succeeded, i.e.,
$$T_{\rm cutoff}:=T_{\rm cohere} - T_{\rm her} - T_{\rm app}.$$
Here, for simplicity, we can assume here that $T_{\rm app}=0$. Note that if $T_{\rm cohere}$ is sufficiently large compared to $T_i$, then one entanglement once established can just wait for the other to become available to swap, which takes
$$T_{12}:=\max\{T_1,T_2\}.$$
On the other hand, when  $T_{\rm cohere}$ is small in relation to $T_{12}$, then $T_{\rm cutoff}$ can be so small that an entanglement can decohere and thus must be restarted (possibly multiple times) before the other one becomes available. Based on this, we  select the generation phase duration and  time slot to be
\[
T_{\rm gen} = \min\big\{\mathbb{E}[T_{12}],  T_{\rm cutoff} \big\}, \quad T_s = T_{\rm gen} + T_{\rm her}.
\]
Since the link entanglements $E_i$ can be well approximated by a Binomial RV, we can approximate $T_i$ by an exponentially distributed RV with rate $\frac{1}{R_i}$ and find
\[
\mathbb{E}[T_{12}] = \mathbb{E}[T_{1}]+\mathbb{E}[T_{2}] - \frac{\mathbb{E}[T_{1}]\mathbb{E}[T_{2}]}{\mathbb{E}[T_{1}] + \mathbb{E}[T_{2}]} \approx \frac{1}{R_1}+\frac{1}{R_2} - \frac{1}{R_1 + R_2}.
\]
Once $T_s$ is selected, we can find the estimated capacity of each link $i$ within a time slot to be
\[
\tilde{c}_i = A_i\times T_s, \quad i=1,2,
\]
where we recall that $A_i$ is the attempt rate of link $i$ (i.e., number of attempts per second). Note that since our model only accepts $c_i$ as an integer, a rounding step is needed. To resolve this, we resort to an MC simulation approach, assigning $r_i = \tilde{c}_i - \lfloor \tilde{c}_i\rfloor$ and
\[
c_i =
\begin{cases}
\lfloor \tilde{c}_i\rfloor   \quad&\text{w.p.}~~~1-r_i\\
\lfloor \tilde{c}_i\rfloor+1 \quad&\text{w.p.}~~~r_i.
\end{cases}
\]
This MC approach is a small extension to the approach in \cite{mai2025optimalorders}, where a simple rounding step was used instead.
As a result, we can proceed to find the throughput using \eqref{thruput} for each sample of $\{c_i\}$, and take the average over a number of samples to be the estimated throughput of our model. Finally, the model predictions of the throughput shown in Figure~5 of the main paper are obtained from using 5000 samples.

\section{Use Case 3: Path Multiplexing}
\label{sec:usecase3}

This use case demonstrates multi-flow routing and link multiplexing on a non-trivial graph with shared trunks. The simulation reproduces (in simplified form) the multiplexing study of a 13-node network with up to five simultaneous flows presented in \cite{aparicio2011multiplexing}.
The goal is to compare \emph{Statistical} vs. \emph{Buffer-Space} multiplexing under controlled contention, and validate the behavior of our implementation.

Compared with the referenced study, our simulation evaluates only two multiplexing schemes—Statistical and Buffer-Space—excluding the Time Division Multiplexing (TDM) variant. Furthermore, we did not incorporate EPR purification to a target fidelity threshold; while purification would reduce raw entanglement rates across all flows, it would not alter the relative comparison between the multiplexing schemes.

\begin{figure}
\centerline{\includegraphics[width=20pc]{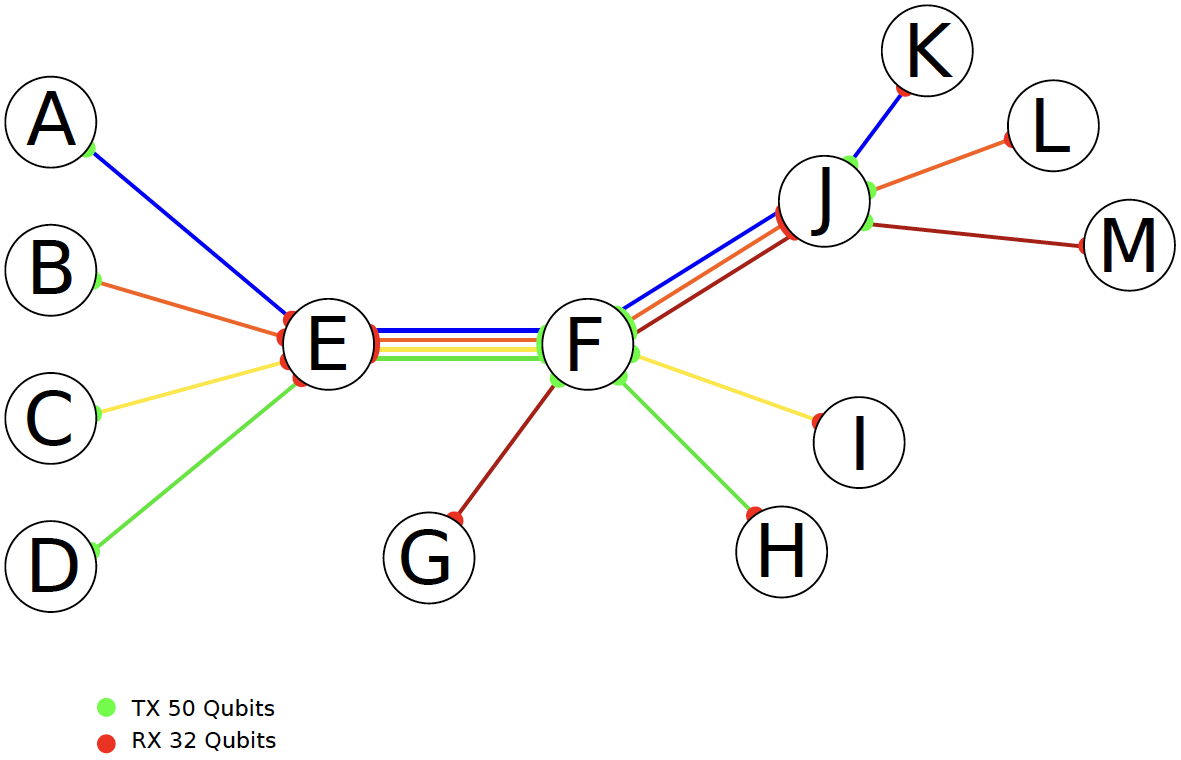}}
\caption{Simulated network. Each color represents one communication flow \cite{aparicio2011multiplexing}}
\label{fig:topology_13}
\end{figure}

\paragraph*{\textbf{Topology and routes}}
The simulated network is shown in Figure \ref{fig:topology_13}.
It consists of a 13-node topology with two shared trunks, $E{\text-}F$ and $F{\text-}J$, and uniform quantum/classical links of 30 km. The left spokes ($A,B,C,D$) meet at $E$, the center trunk is $E{\text-}F{\text-}J$, the right spokes ($K,L,M$) branch from $J$, and $G,H,I$ attach around $F$.

All quantum links are length 30 \,km with loss $\alpha=0.17$\,dB/km; transmitter/receiver memory capacities are 50/32 qubits, respectively. The five end-to-end \emph{flows} and their static routes are:
\[
\begin{aligned}
\mathrm{AK}:&\; A\!\to\!E\!\to\!F\!\to\!J\!\to\!K,\qquad
\mathrm{BL}: B\!\to\!E\!\to\!F\!\to\!J\!\to\!L,\\
\mathrm{CI}:&\; C\!\to\!E\!\to\!F\!\to\!I,\qquad\qquad\ \;
\mathrm{DH}: D\!\to\!E\!\to\!F\!\to\!H,\\
\mathrm{GM}:&\; G\!\to\!F\!\to\!J\!\to\!M.
\end{aligned}
\]

\paragraph*{\textbf{Protocol settings}}
We fix base Bell-pair fidelity at $0.99$ (per-link initialization), enable swapping with probability $p_{\mathrm{swap}}=0.5$ using an ASAP policy, and set coherence time $t_{\mathrm{cohere}}=100$\,ms. Detector and source efficiencies were set at $\eta_d{=}0.5$, $\eta_s{=}0.8$ respectively.

\paragraph*{\textbf{Scenarios}}
We evaluate the five standard scenarios used in the reference work to control how many flows contend at $E{\text-}F$ and $F{\text-}J$:
\begin{enumerate}
  \item Baselines (\texttt{AK}, \texttt{BL}, \texttt{CI}, \texttt{DH}, \texttt{GM} alone),
  \item \texttt{1) AK+CI} (one contested link $E{\text-}F$),
  \item \texttt{2) AK+BL} (two contested links $E{\text-}F$ and $F{\text-}J$),
  \item \texttt{3) AK+CI+DH} (two competitors share $E{\text-}F$ with AK),
  \item \texttt{4) AK+BL+CI+DH+GM} (all five flows; both trunks contested).
\end{enumerate}

\paragraph*{\textbf{Multiplexing strategies}}
The simulation compares two schemes:
\begin{itemize}
  \item Statistical multiplexing. No prior reservation or static split; swapping opportunities on contested links are consumed on a best-effort basis.
  \item Buffer-Space multiplexing. Per-hop Multiplexing Vectors (MVs) statically split left-right memory on contested links according to the active flow set. Uncontested links use full left-right memory capacity.
\end{itemize}
On the two contested trunks, the buffer-space splits are as follows:
\begin{enumerate}
  \item \emph{$E{\text-}F$ under AK+CI:} $E{:}16$, $F{:}25$ per flow.
  \item \emph{$E{\text-}F$ under AK+BL:} $E{:}16$, $F{:}25$ per flow (and $F{\text-}J$ is also contested).
  \item \emph{$E{\text-}F$ under AK+CI+DH:} $E{:}(11,11,10)$ and $F{:}(17,17,16)$ for (AK, CI, DH).
  \item \emph{All five flows:} $E{\text-}F$ has four flows (AK, BL, CI, DH) with $E{:}8$ each; $F{:}(17,17,16,12)$ for (AK, BL, CI, DH). For $F{\text-}J$, three flows (AK, BL, GM) share $F{:}(17,17,16)$ and $J{:}(11,11,10)$.
\end{enumerate}
These allocations match the reference study's buffer-space policy and create explicit, non-overlapping memory bands on the congested links.

\begin{figure}
\centerline{\includegraphics[width=20pc]{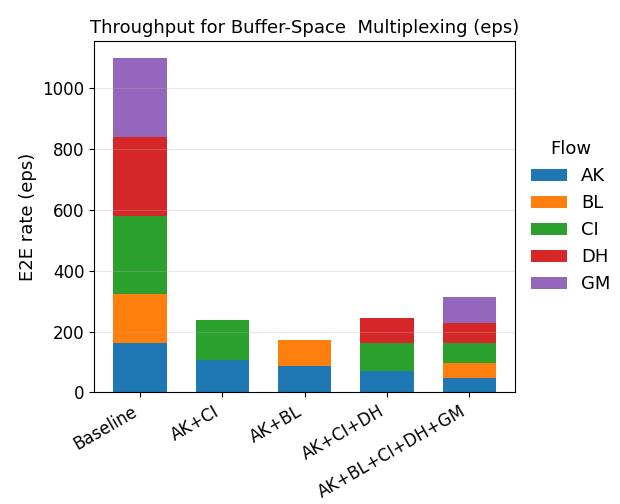}}
\caption{Throughput of buffer space multiplexing compared to uncontested flows (baseline)}
\label{fig:buffer_space_rate}
\end{figure}

\begin{figure}
\centerline{\includegraphics[width=20pc]{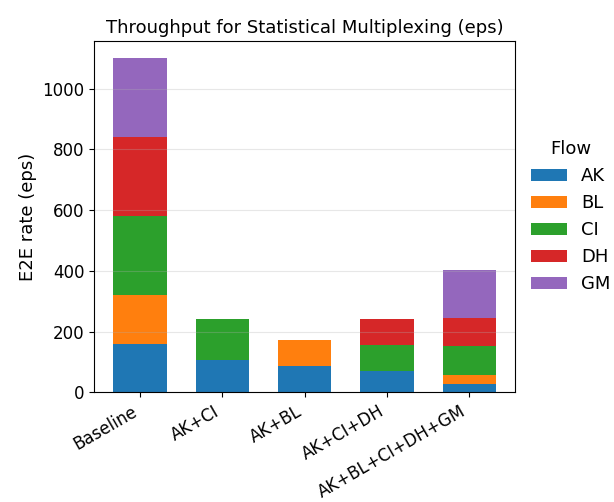}}
\caption{Throughput of statistical multiplexing compared to uncontested flows (baseline)}
\label{fig:statistical_rate}
\end{figure}

\paragraph*{\textbf{Results}}
For each strategy $\times$ scenario, we execute 100 runs (distinct seeds), then record per-flow (source-side) end-to-end rate in entanglement per second (eps).
The rates were aggregated and rendered in Figures \ref{fig:buffer_space_rate} and \ref{fig:statistical_rate}.
The leftmost bar shows the sum of per-flow baselines (each flow run alone), and the remaining bars show the realized per-flow contributions under each flow combination scenario.

Despite these simplifications, the trends reported match those of the replicated study. Statistical multiplexing efficiently exploits idle resources along uncontested subpaths and typically yields higher total throughput, particularly when all five flows run concurrently. Buffer-space multiplexing offers predictability and isolation, but may underutilize memory on contested trunks. Fairness across flows remains high for both multiplexing schemes.

\paragraph*{\textbf{Discussion}}
This use case also highlights several open research questions about the practical design of multiplexing protocols in quantum networks. Our experiments confirmed the intuitive expectation that statistical multiplexing should outperform buffer-space multiplexing in terms of aggregate throughput.
However, the realistic design of MQNS also revealed that this advantage is only realized when the swapping decisions along a path are coordinated.
For example, consider a situation where node~$E$ connects an $E{\text-}F$ entanglement with $D{\text-}E$, while almost simultaneously node~$F$ connects the same $E{\text-}F$ entanglement with $F{\text-}I$. Under the \texttt{SWAP-ASAP} (or other simultaneous-swapping) policy, the resulting entanglement becomes $D{\text-}I$, which does not correspond to any configured flow in the network. We refer to such cases as \textit{conflictual swapping decisions}.
In simulation, one can work around this issue by coordinating swap operations artificially (e.g., through global scheduling or instantaneous swap updates), thereby revealing the performance ceiling of statistical multiplexing. But this points to an architectural challenge: how should realistic protocols coordinate swaps across multiple flows in a distributed environment with latency and signaling delays?

This raises several research questions:
\begin{itemize}
  \item What minimal level of classical coordination is required to ensure that statistical multiplexing achieves its throughput advantage in practice?
  \item How should memory access be abstracted and exposed to upper layers so that conflicting swap attempts can be detected and avoided?
  \item What is the impact of realistic signaling delays, message
  processing times, and classical control overhead on the relative performance of multiplexing strategies?
\end{itemize}

By raising these questions, MQNS serves as both a benchmarking platform and a design probe. It helps identify where new protocol mechanisms, interfaces, and signaling are required to make entanglement distribution both efficient and realistic in quantum networks.

\section{Use Case 4: Simulator Scalability}

To evaluate the scalability of MQNS, we measured the wall-clock execution time over randomly generated quantum network topologies with increasing size and channel capacity, and we ran the same simulation scenarios on SeQUeNCe under identical conditions.

It is important to emphasize that this comparison is not intended to claim superiority over SeQUeNCe.
Indeed, SeQUeNCe can operate at a finer physical granularity by modeling device-level characteristics and optionally tracking full density matrices.
Instead, the purpose of this use case is to demonstrate the scalability advantage of avoiding explicit photon-level simulation in large networks, while still achieving closely matching entanglement distribution behavior, as validated earlier in Use Case 1 of the main paper.

A random network topology was generated for each configuration, with the number of edges chosen such that the average node degree was approximately 2.5.
The network sizes tested ranged from 16 to 512 nodes.
Quantum channels were assigned a memory capacity of 10, 50, and 100 qubits.

Entanglement generation was simulated under the following conditions:
\begin{itemize}
 \item Fiber attenuation: 0.2 dB/km
 \item Detector efficiency: 0.95
 \item Source efficiency: 0.95
 \item Memory coherence time: 5 ms
\end{itemize}

Both simulators executed the same end-to-end entanglement workload, for a simulation time of 3 seconds.
Proactive forwarding with statistical multiplexing was used, and entanglement swapping was performed using the SWAP-ASAP policy.
For each network size, the number of entanglement requests was proportional to the network size, with approximately 20\% of the nodes selected as source-destination pairs.
The paths were pre-installed using hop-count Dijkstra routing.

We ran all simulations in Docker containers on a FABRIC testbed~\cite{fabric-2019} virtual machine.
Each simulator container was pinned to one dedicated EPYC 7H12 processor core at 2.6 GHz frequency, isolated from other usage.

\begin{figure}[t]
\centering
\includegraphics[width=18.5pc]{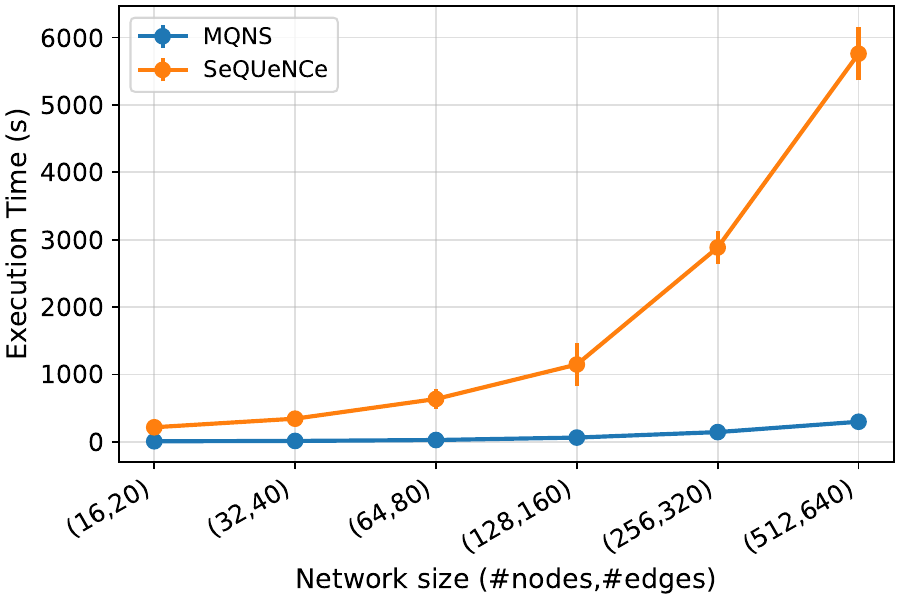}
\caption{Simulation execution time for increasing network sizes and 10 qubits channel capacity.}
\label{fig:sca10}

\vspace{1em} 

\includegraphics[width=18.5pc]{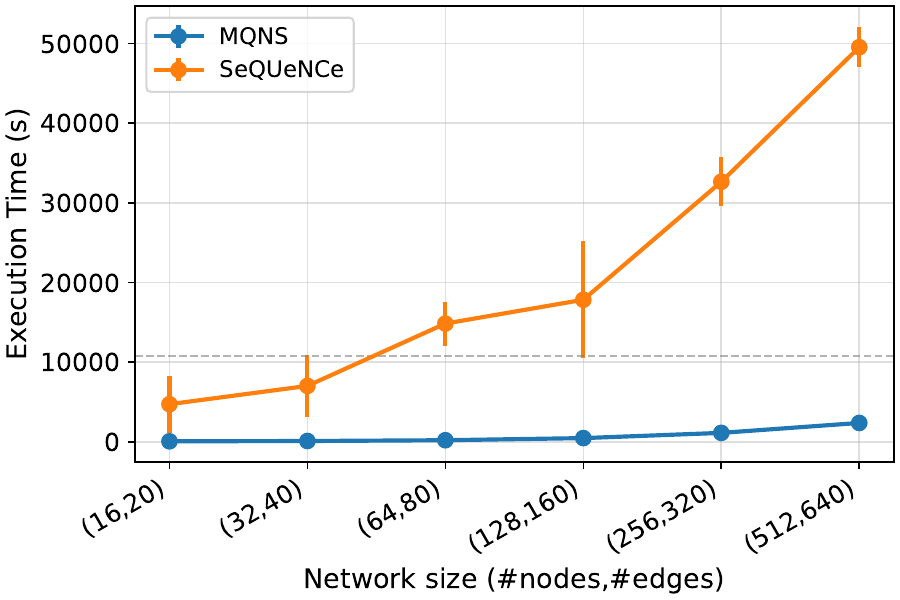}
\caption{Simulation execution time for increasing network sizes and 50 qubits channel capacity.}
\label{fig:sca50}

\vspace{1em}

\includegraphics[width=18.5pc]{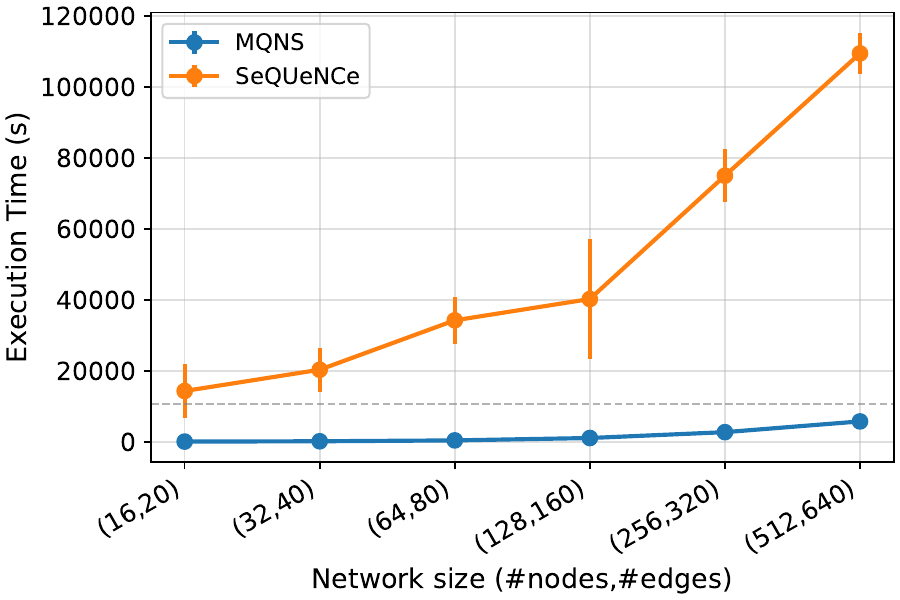}
\caption{Simulation execution time for increasing network sizes and 100 qubits channel capacity.}
\label{fig:sca100}
\end{figure}

Figures \ref{fig:sca10}, \ref{fig:sca50}, and \ref{fig:sca100} report the execution-time scaling behavior for channel capacities of 10, 50, and 100 qubits, respectively, averaged over 5 runs per scenario.
For SeQUeNCe, if a run was still ongoing after \num{10800} seconds (3 hours) of wall-clock time, it would be forcibly terminated, and the execution-time was extrapolated to 3-second simulation time based on the simulation timeline progress upon termination.

As expected, execution time increases with network size for both simulators.
However, MQNS scales substantially more efficiently, completing all scenarios up to 512 nodes well within practical execution times on commodity hardware.
SeQUeNCe becomes progressively slower and fails to finish several runs within the \qty{10800}{s} cutoff — especially at higher channel capacities where MQNS remains tractable.

\section*{Reproducibility}
All the use cases presented in the main manuscript and this supplemental material are available in the \textit{examples} folder of the simulator's repository.

\bibliographystyle{IEEEtran}
\bibliography{ref.bib}

\end{document}